\documentclass[a4paper,11pt]{article}
\usepackage{pos}
 \usepackage[nice]{nicefrac}
\def\HP{\hphantom{\alpha}} % horizontal

\def\be{\begin{equation}}
	\def\ee{\end{equation}}
\newcommand{\bel}[1]{\begin{eqnarray}\label{#1}}
	\newcommand{\eel}{\end{eqnarray}}
\def\barr{\begin{array}}
	\def\earr{\end{array}}
\def\beq{\begin{eqnarray}}
	\def\eeq{\end{eqnarray}}
\def\bfig{\begin{figure}}
	\def\efig{\end{figure}}

\def\CHI{\chi}
\newcommand{\nn}{\nonumber}
\newcommand{\f}[2]{\frac{#1}{#2}}
\newcommand{\onehalf}{{\nicefrac{1}{2}}}

\newcommand{\rf}[1]{Eq.~(\ref{#1})}

\def\a{\alpha}
\def\b{\beta}
\def\g{\gamma}
\def\d{\delta}

\def\LR{\left(} % round
\def\RR{\right)}
\def\HP{\hphantom{\alpha}} % horizontal

\newcommand{\sh}[1]{\sinh#1}
\newcommand{\ch}[1]{\cosh#1}

\def\half{\frac{1}{2}}

\newcommand{\lab}[1]{\label{#1}}
\def\nn{\nonumber}

\def\half{\frac{1}{2}}

\def\n0{n_{(0)}}
\def\e0{\varepsilon_{(0)}}
\def\P0{P_{(0)}}
\title{Formalism of hydrodynamics with spin degrees of freedom}
\author*{Rajeev Singh}
\affiliation{Institute of Nuclear Physics Polish Academy of Sciences,\\ PL-31-342 Krak\'ow, Poland}
\emailAdd{rajeev.singh@ifj.edu.pl}
\abstract{In this article we review the perfect-fluid hydrodynamics with spin framework proposed recently. This framework generalises the standard relativistic hydrodynamics framework to include spin degrees of freedom and provides a natural method to describe the spin polarization evolution of massive spin 1/2 particles. This formalism is based on the GLW (de Groot - van Leeuwen - van Weert) energy-momentum tensor and spin tensor. We show here using Bjorken model that how this spin hydrodynamics framework may be used for the determination of the observables which describes the particle polarization measured in the experiment.}

\FullConference{%
  The Eighth Annual Conference on Large Hadron Collider Physics-LHCP2020 \\
  25-30 May, 2020\\
  online}

\begin{document}
\maketitle
\section{Introduction} 
Measurements of $\Lambda$ hyperons spin polarization made by the STAR Collaboration~\cite{STAR:2017ckg,Adam:2018ivw,Adam:2019srw,Niida:2018hfw} created immense interest in the theoretical physics community focusing on the interplay between the orbital angular momentum of the matter and particle polarization in the relativistic heavy-ion collisions~\cite{Becattini:2009wh,Becattini:2013fla,Montenegro:2017rbu,Montenegro:2017lvf,Becattini:2018duy,Boldizsar:2018akg,Prokhorov:2018bql,Florkowski:2019voj,Weickgenannt:2019dks,Hattori:2019lfp,Ambrus:2019ayb,Sheng:2019kmk,Prokhorov:2019cik,Ivanov:2019wzg,Hattori:2019ahi,Xie:2019jun,Liu:2019krs,Wu:2019eyi,Becattini:2019ntv,Zhang:2019xya,Li:2019qkf,Florkowski:2019gio,Deng:2020ygd,Fukushima:2020qta,Prokhorov:2019yft,Yang:2018lew,Liu:2020ymh,Tabatabaee:2020efb,Bhadury:2020puc,Liu:2020flb,Yang:2020hri}. Thermal-based models describing the global spin polarization~\cite{Becattini:2016gvu,Karpenko:2016jyx,Li:2017slc,Xie:2017upb,Taya:2020sej,Becattini:2020ngo}, somehow fail to explain the differential results~\cite{Adam:2019srw}. These models deal with the spin polarization at the freeze-out where they assume that thermal vorticity determines spin polarization of particles~\cite{Becattini:2007sr,Becattini:2013fla}. Following Refs.~\cite{Florkowski:2017ruc,Florkowski:2017dyn,Florkowski:2018fap,Florkowski:2018ahw}, we investigate the spacetime evolution of particle spin polarization by including the spin degrees of freedom in the standard hydrodynamic formalism in the Bjorken hydrodynamical setup~\cite{Florkowski:2019qdp}. 
\section{Hydro-dynamical equations and spin polarization tensor}
For particles with spin-$\onehalf$, perfect fluid hydrodynamics formalism is formed on the conservation laws for charge, energy and linear momentum and, angular momentum, which is based on GLW~\cite{DeGroot:1980dk} energy-momentum tensor, $T^{\a\b}_{\rm GLW}$, and spin  tensor, $S^{\a\b\g}_{\rm GLW}$\footnote{Herein we assume small spin polarization of particles ($|\omega_{\mu\nu}| < 1$).}, namely~\cite{Florkowski:2017ruc,Florkowski:2017dyn,Florkowski:2018fap}
\begin{eqnarray} 
\quad\partial_\mu N^\mu = 0,  \qquad
\partial_\mu T^{\mu\nu}_{\rm GLW} = 0, \qquad
\partial_\lambda  S_{\rm GLW}^{\lambda, \alpha \beta} =T_{\rm GLW}^{\beta \alpha} -T_{\rm GLW}^{\alpha \beta}, 
\label{eom}
\end{eqnarray}
with
\vspace{-0.3cm}
\bel{Tmn}
N^\alpha = n U^\alpha, \qquad T^{\a\b}_{\rm GLW} = (\varepsilon + P ) U^\a U^\b - P g^{\a\b},
\eel
with $N^\alpha$ being the net baryon charge current and $\varepsilon$, $P$, $n$ is the energy density, pressure and baryon number density, respectively, and $U^\beta$ is the time-like hydrodynamic flow four-vector. Due to the symmetricity of energy-momentum tensor in Eq.~(\ref{eom}), conservation of the angular momentum in-turn implies conservation of the spin part separately~\cite{Florkowski:2018ahw}. The spin tensor~\cite{Florkowski:2017dyn} is $S^{\alpha , \beta \gamma }_{\rm GLW}
=  {\cal C} \left[ n_{(0)}(T) U^\alpha \omega^{\beta\gamma}  +  {\cal A}_{(0)} \, U^\a U^\d U^{[\b} \omega^{\g]}_{\HP\d} + \, {\cal B}_{(0)} \, \Big( 
U^{[\b} \Delta^{\a\d} \omega^{\g]}_{\HP\d}
+ U^\a \Delta^{\d[\b} \omega^{\g]}_{\HP\d}
+ U^\d \Delta^{\a[\b} \omega^{\g]}_{\HP\d}\Big) \right]$\\
where ${\cal C}=\ch(\xi)$, ${\cal B}_{(0)} =-\frac{2}{\hat{m}^2} s_{(0)}(T)$ and $ 
{\cal A}_{(0)}  = -3{\cal B}_{(0)} +2 n_{(0)}(T)$, 
and $n_{(0)}(T)$, $s_{(0)}(T)$ is the number density and entropy density of spin-less and neutral massive Boltzmann particles~\cite{Florkowski:2010zz}, respectively, with $\Delta^{\a\b}$ being the spatial projection operator orthogonal to $U$ with $\xi=\mu/T$ (i.e. baryon chemical potential over temperature) and  $\hat{m}=m/T$, $m$ is the particle mass.\\
Being an asymmetric second rank tensor, the polarization tensor $\omega_{\a\b}$ can be written as
\beq
\omega_{\a\b} &=& \kappa_\a U_\b - \kappa_\b U_\a + \epsilon_{\a\b\g\d} U^\g \omega^{\d}, \lab{spinpol1}
\eeq
where $\kappa$ and $\omega$ are four-vectors orthogonal to fluid flow four-vector $U$. 
For system with boost-invariant flow and transversely homogeneous set-up the following basis are useful for further calculations
\begin{eqnarray} 
U^\a = \LR \ch(\eta), 0,0, \sh(\eta) \RR, \quad X^\a = \LR 0, 1,0, 0 \RR,\nn \\
Y^\a = \LR 0, 0,1, 0 \RR, \quad Z^\a = \LR \sh(\eta), 0,0, \ch(\eta) \RR
\lab{BIbasis}
\end{eqnarray}
with $\eta = \half \ln((t+z)/(t-z))$ being the space-time rapidity. Using these four-vector basis, $\kappa^{\a}$ and $\omega^{\a}$ can be decomposed in the following way where all the scalar coefficients are function of proper time $(\tau)$ only.
\begin{eqnarray} 
\kappa^\a =  {C}_{\kappa X} (\tau) X^\a + {C}_{\kappa Y}(\tau) Y^\a + {C}_{\kappa Z}(\tau) Z^\a,\quad \omega^\a =  {C}_{\omega X}(\tau) X^\a + {C}_{\omega Y}(\tau) Y^\a + {C}_{\omega Z}(\tau) Z^\a \lab{eq:o_decom}
\end{eqnarray}
Using \rf{eq:o_decom} in spin conservation law and projecting the resulting tensorial equation on $U_\b X_\g$, $U_\b Y_\g$, $U_\b Z_\g$, $Y_\b Z_\g$, $X_\b Z_\g$, and, $X_\b Y_\g$ we get equations of motion for scalar coefficients which turn out to evolve independently. Due to the rotational symmetry in the transverse plane, coefficients ${C}_{\kappa X}$, ${C}_{\kappa Y}$ and ${C}_{\omega X}$, ${C}_{\omega Y}$ follow the same structure of differential equations, respectively. 
%%%%%%%%%%%%%%
\section{Particle spin polarization at freeze-out}
Spin polarization tensor evolution permit us to calculate the mean spin polarization per particle defined as $\langle\pi_{\mu}\rangle=E_p\frac{d\Pi _{\mu }(p)}{d^3 p}/E_p\frac{d{\cal{N}}(p)}{d^3 p}$~\cite{Florkowski:2018ahw} with
\beq
E_p\frac{d\Pi _{\mu }(p)}{d^3 p} &=& -\f{ \cosh(\xi)}{(2 \pi )^3 m}
\int
\Delta \Sigma _{\lambda } p^{\lambda } \,
e^{-\beta \cdot p} \,
\tilde{\omega }_{\mu \beta }p^{\beta },\nn \\ E_p\frac{d{\cal{N}}(p)}{d^3 p}&=&
\f{4 \cosh(\xi)}{(2 \pi )^3}
\int
\Delta \Sigma _{\lambda } p^{\lambda } 
\,
e^{-\beta \cdot p},
\eeq
where $E_p\frac{d\Pi _{\mu }(p)}{d^3 p}$ being the total value of the Pauli-Luba\'nski vector for particles and $E_p\frac{d{\cal{N}}(p)}{d^3 p}$ being the momentum density of all particles having momentum $p$. Using canonical boost, the polarization vector $\langle\pi^{\star}_{\mu}\rangle$ in the particle rest frame can be obtained, where its longitudinal component is~\cite{Florkowski:2019qdp}
\begin{eqnarray}
\langle\pi^{\star}_{z}\rangle&=&\frac{1}{8m}
\Big[\left(\frac{m\ch(y_p)+m_T}{m_T \ch(y_p)+m}\right)\left[\CHI\left({C}_{\kappa X} p_y-{C}_{\kappa Y} p_x\right)+2 {C}_{\omega Z} m_T  \right]\nn\\
&& +\frac{\CHI \,m\,\sh(y_p) \left({C}_{\omega X} p_x+{C}_{\omega Y} p_y\right)}{m_T \ch(y_p)+m}\Big]
\end{eqnarray}
where $\CHI\left( \hat{m}_T \right)=\left( K_{0}\left( \hat{m}_T \right)+K_{2}\left( \hat{m}_T \right)\right)/K_{1}\left( \hat{m}_T \right)$, $\hat{m}_T=m_T/T$ and $y_p$ is the rapidity.
\section{Results and Conclusion}
\begin{figure}
\centering
 \includegraphics[width=0.32\textwidth]{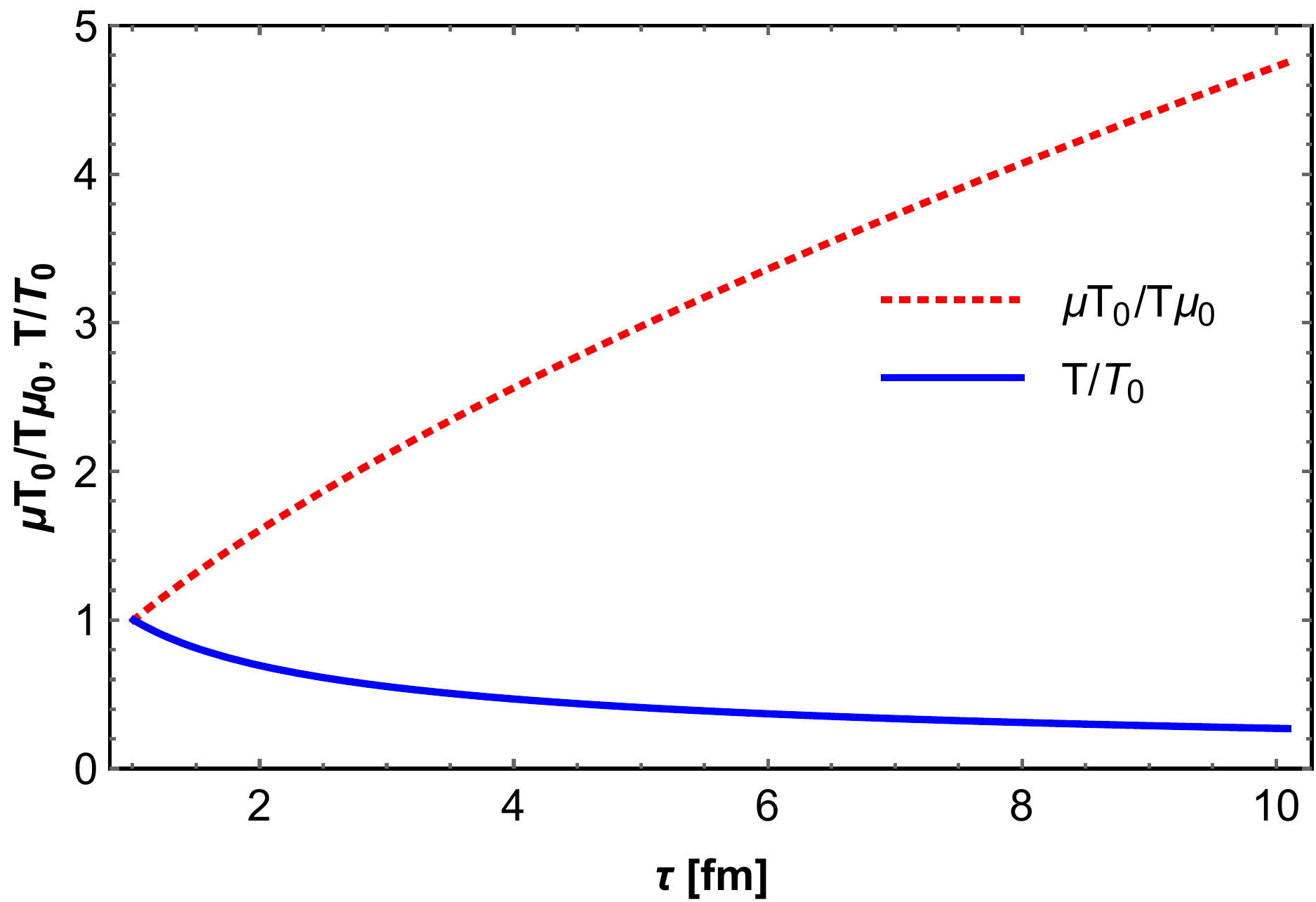}
 \includegraphics[width=0.32\textwidth]{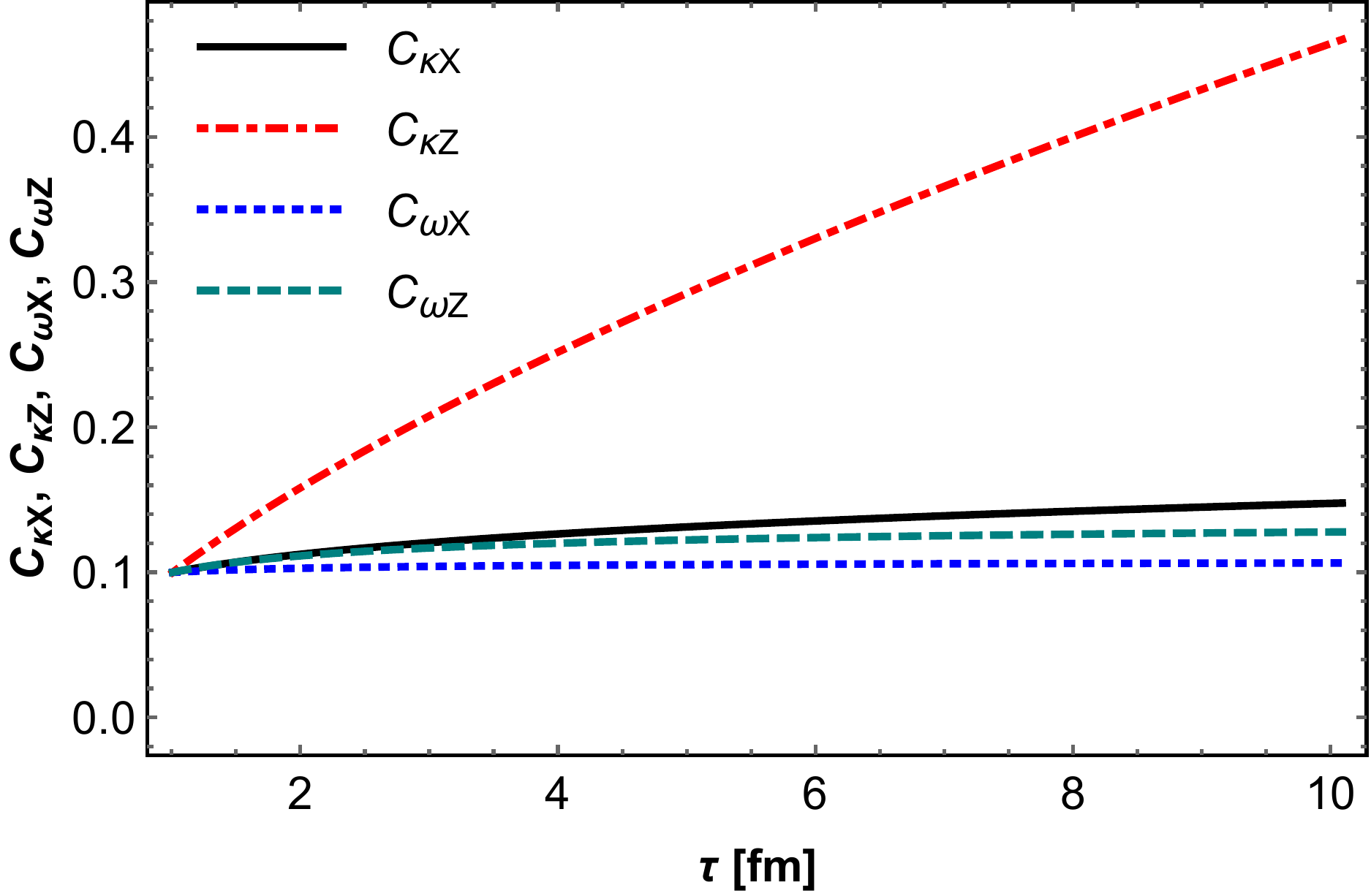}\\
 \includegraphics[width=0.32\textwidth]{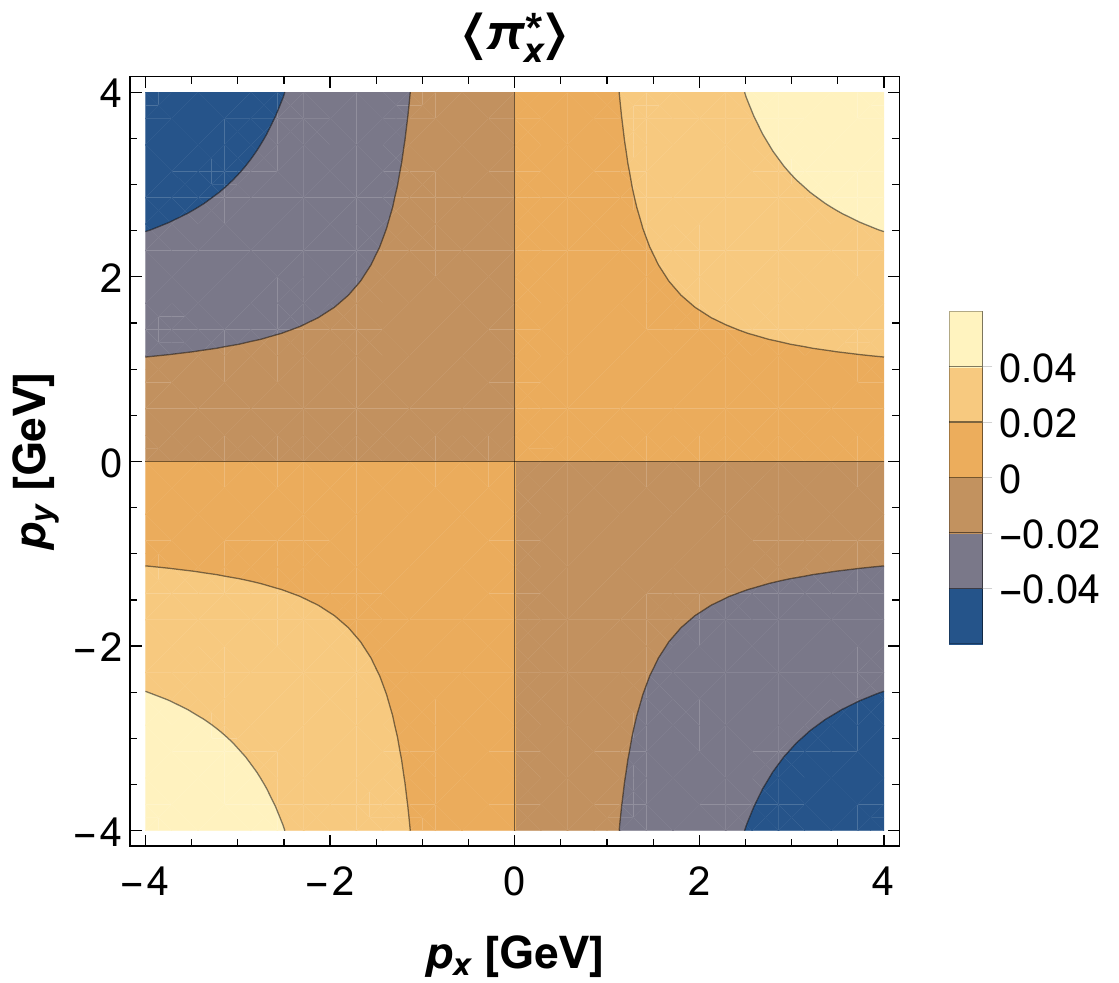}
 \includegraphics[width=0.32\textwidth]{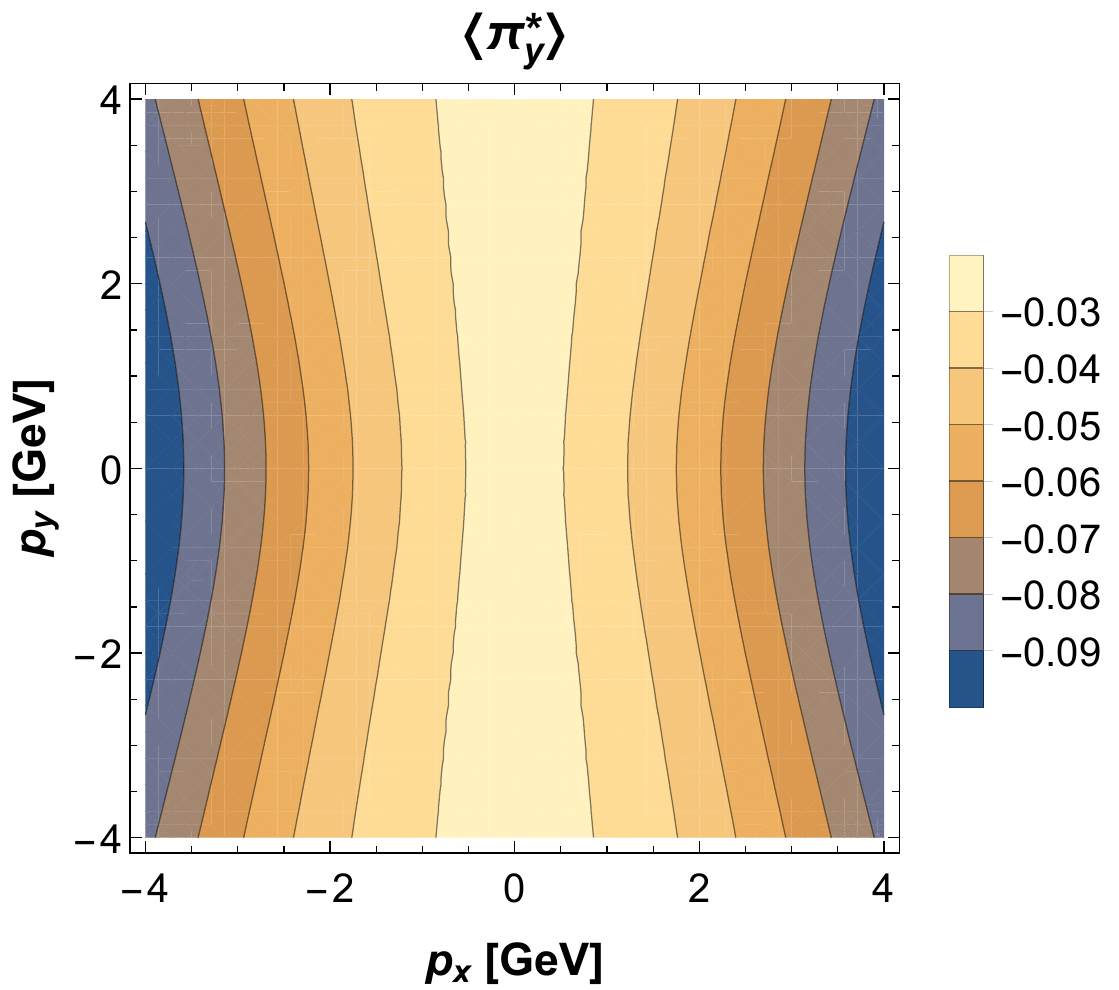}
 \includegraphics[width=0.32\textwidth]{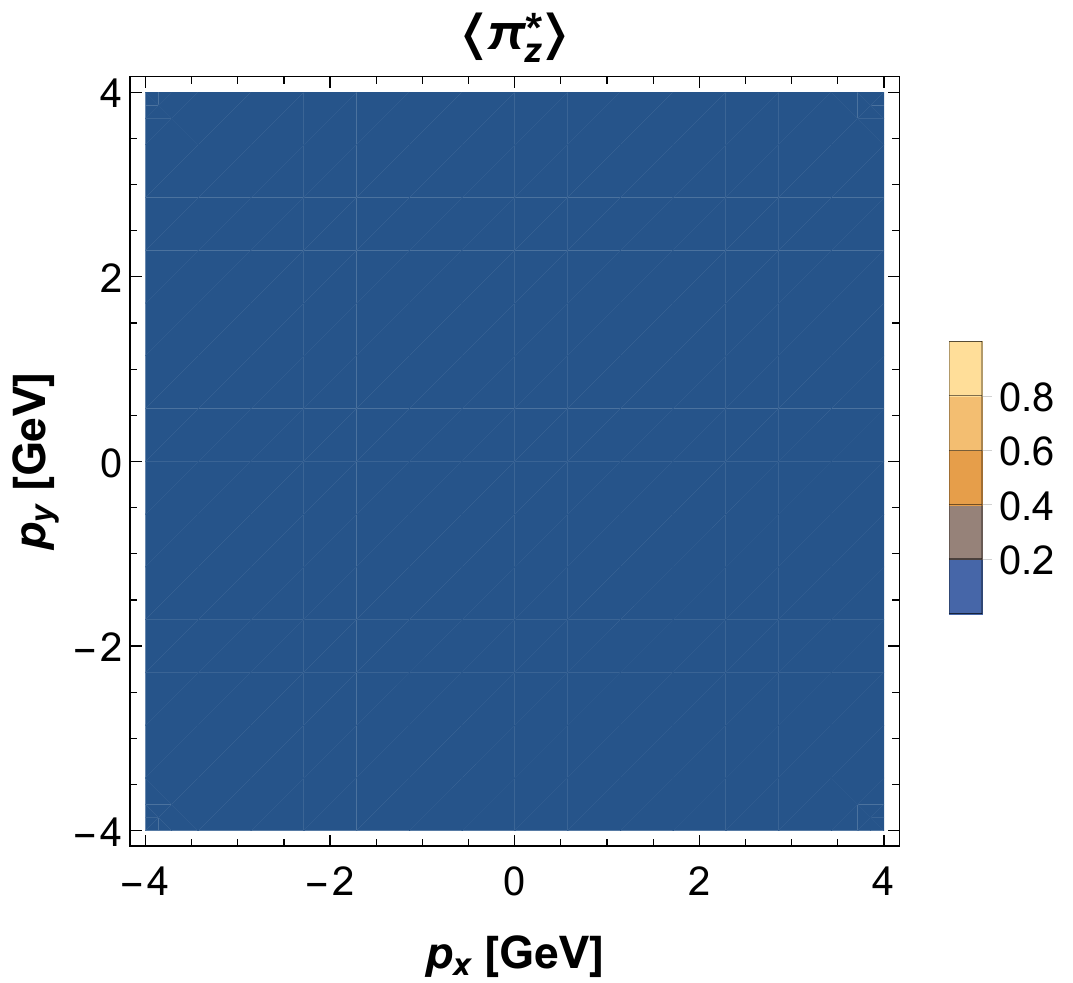}
\caption{Above panel left: Proper-time dependence of temperature $T$ rescaled by its initial value $T_0$ (solid line) and the ratio of $\mu$ (baryon chemical potential) and $T$ (temperature) rescaled by the initial ratio $\mu_0/T_0$ (dotted). Above panel right: Proper-time dependence of the coefficients ${C}_{\kappa X}$, ${C}_{\kappa Z}$, ${C}_{\omega X}$ and ${C}_{\omega Z}$. Below panel: Components of the PRF average polarization three-vector of $\Lambda$'s with the initial conditions $\mu_0=800$~MeV,
$T_0=155$~MeV, $C_{\kappa, 0}=(0,0,0)$ and $ C_{\omega, 0}=(0,0.1,0)$ for mid-rapidity.}
\label{fig:FH}
\end{figure}

For the Bjorken flow, charge current conservation and energy and linear momentum conservation is written as $\frac{dn}{d\tau}+\frac{n}{\tau}=0$ and $\frac{d\varepsilon}{d\tau}+\frac{(\varepsilon+P)}{\tau}=0$, respectively.
In Fig.\ref{fig:FH} (above panel left), proper-time dependence of temperature and baryon chemical potential is shown, where temperature decreases with proper-time, whereas ratio of baryon chemical potential over temperature increases with proper time. In Fig.\ref{fig:FH} (above panel right), dependence of the $C$ coefficients on proper time is shown. Using thermodynamical quantities and $C$ coefficients at freeze-out, different components of the PRF average polarization vector $\langle\pi^{\star}_{\mu}\rangle$ are obtained, see Fig.\ref{fig:FH} (below panel). We notice that  $\langle\pi^{\star}_{y}\rangle$ is negative which reflects the initial system's spin angular momentum. At midrapidity, the longitudinal component of polarization vector ($\langle\pi^{\star}_{z}\rangle$) is zero and $\langle\pi^{\star}_{x}\rangle$ shows quadrupole structure.
Our results presented here does not show quadrupole structure of the longitudinal polarization observed in experiments because of the symmetries we assumed. We presented results using the perfect-fluid hydrodynamics framework with spin for the Bjorken scenario~\cite{Bjorken:1982qr,Rezzolla:2013zz}. Our approach show that scalar coefficients $C$ which describes spin polarization evolve independently with weak proper-time dependence. Our results may be used to determine the particle spin polarization at freeze-out and we have shown that particle spin polarization formed at freeze-out hypersurface depicts the initial polarization direction.\\
I thank Wojciech Florkowski and Radoslaw Ryblewski for clarifying and inspiring discussions. This research was supported in part by the Polish National Science Center Grants No. 2016/23/B/ST2/00717 and No. 2018/30/E/ST2/00432.
%%%%%%

\end{document}